\documentclass[aps, prl, amsmath,amssymb,twocolumn, groupedaddress, superscriptaddress, reprint]{revtex4-1}
\usepackage{graphicx}
\usepackage{dcolumn}
\usepackage[utf8]{inputenc}
\usepackage[dvipsnames]{xcolor}
\usepackage[colorlinks,linkcolor=blue,citecolor=blue,urlcolor=blue]{hyperref}
\usepackage{stix}
\renewcommand{\vec}[1]{\mathbf{#1}}
\newcommand{\gvec}[1]{\boldsymbol{#1}}
\usepackage[dvipsnames]{xcolor}
\usepackage{xspace}

\begin{document}

\title{Coherent Modulation of Quasiparticle Scattering Rates in a Photoexcited Charge-Density-Wave System} 

\author{J. Maklar}
\affiliation{Fritz-Haber-Institut der Max-Planck-Gesellschaft, Faradayweg 4-6, 14195 Berlin, Germany}
\author{M. Schüler}
\affiliation{Stanford Institute for Materials and Energy Sciences, SLAC National Accelerator Laboratory, 2575 Sand Hill Road, Menlo Park, CA 94025, USA}
\author{Y. W. Windsor}
\author{C. W. Nicholson}
\altaffiliation[Current address: ]{Department of Physics and Fribourg Center for Nanomaterials, University of Fribourg, Chemin du Musée 3, CH-1700 Fribourg, Switzerland}
\author{M. Puppin}
\altaffiliation[Current address: ]{Laboratory of Ultrafast Spectroscopy, ISIC, Ecole Polytechnique Fédérale de Lausanne (EPFL), CH-1015 Lausanne, Switzerland}
\affiliation{Fritz-Haber-Institut der Max-Planck-Gesellschaft, Faradayweg 4-6, 14195 Berlin, Germany}
\author{P. Walmsley}
\author{I. R. Fisher}
\affiliation{Stanford Institute for Materials and Energy Sciences, SLAC National Accelerator Laboratory, 2575 Sand Hill Road, Menlo Park, CA 94025, USA}
\affiliation{Geballe Laboratory for Advanced Materials and Department of Applied Physics, Stanford University, Stanford, CA 94305, USA}
\author{M. Wolf}
\affiliation{Fritz-Haber-Institut der Max-Planck-Gesellschaft, Faradayweg 4-6, 14195 Berlin, Germany}
\author{R. Ernstorfer}
\affiliation{Fritz-Haber-Institut der Max-Planck-Gesellschaft, Faradayweg 4-6, 14195 Berlin, Germany}
\affiliation{Institut für Optik und Atomare Physik, Technische Universität Berlin, Straße des 17. Juni 135, 10623 Berlin, Germany}
\author{M. A. Sentef}
\affiliation{Max Planck Institute for the Structure and Dynamics of Matter, Luruper Chaussee 149, 22761 Hamburg, Germany}
\author{L. Rettig}
\email{rettig@fhi-berlin.mpg.de}
\affiliation{Fritz-Haber-Institut der Max-Planck-Gesellschaft, Faradayweg 4-6, 14195 Berlin, Germany}

\date{\today}

\begin{abstract}
We present a complementary experimental and theoretical investigation of relaxation dynamics in the charge-density-wave (CDW) system TbTe$_3$ after ultrafast optical excitation. Using time- and angle-resolved photoemission spectroscopy, we observe an unusual transient modulation of the relaxation rates of excited photocarriers. A detailed analysis of the electron self-energy based on a nonequilibrium Green's function formalism reveals that the phase space of electron-electron scattering is critically modulated by the photoinduced collective CDW excitation, providing an intuitive microscopic understanding of the observed dynamics and revealing the impact of the electronic band structure on the self-energy.
\end{abstract}

\maketitle

A defining feature of metals is that they feature electronic states around the Fermi level $E_\mathrm{F}$, allowing for low-energy excitations and thus enabling efficient electronic transport. A central quantity that encodes such quasiparticle excitations is the electron self-energy $\Sigma(E,\boldsymbol{k})$, which describes all interactions with the electronic quasiparticle at energy $E$ and momentum $\boldsymbol{k}$. In simple metals, the self-energy is dominated by electron-electron (e--e) scattering and is well-described by Fermi-liquid theory~\cite{bauer2015hot}. However, in more complex materials, an interplay between e--e, electron-phonon (e--ph), electron-impurity scattering, and the electronic band structure, governing the available phase space for particular quasiparticle excitations, has to be considered~\cite{chulkov2006electronic}. Thus, for most materials, a quantitative theoretical treatment proves challenging\cite{Calandra2007_Graphene, King2014_SrTitanates}, and a detailed understanding of the self-energy -- essential for disentangling the intricate interactions in complex solids -- is still missing. 

A key technique to investigate quasiparticle interactions in solids directly on their intrinsic timescales is time- and angle-resolved photoemission spectroscopy (trARPES)~\cite{Bovensiepen2012Sep}, in which a femtosecond (fs) optical pump pulse generates an excited-carrier population. Probing the transient quasiparticle band structure by a time-delayed UV/XUV pulse during the subsequent recovery elucidates charge- and energy redistribution processes and allows disentangling competing contributions to the self-energy~\cite{perfetti2007ultrafast, kirchmann2010quasiparticle, hellmann2012time, gierz2014non, yang2015inequivalence, rameau2016energy, ulstrup2014ultrafast, Pomarico2017}. Additionally, optical excitation allows for transient manipulation of specific material properties with high selectivity by tailoring the excitation wavelength, polarization, pulse energy, and length. Prominent examples of this rapidly growing research field include photostabilization of superconductivity~\cite{fausti_light-induced_2011, mitrano_possible_2016, Budden2021May}, dynamical modification of the band structure due to the formation of photon-dressed states~\cite{wang_observation_2013, mahmood2016selective}, and stabilization of nonequilibrium metastable states~\cite{Rini2007Sep, stojchevska2014ultrafast, Zhang2016Sep}. However, the microscopic processes underlying such transitions and the properties of light-induced states often lack a clear understanding. Intriguingly, time-domain studies offer a natural route to access the transient quasiparticle interactions during the optical control of condensed matter properties, providing an understanding on the microscopic level of the self-energy. As most (photoinduced) phase transitions in complex solids involve an insulator-to-metal transition, it is of special interest how a modification of the low-energy electronic states impacts the electron self-energy and which particular scattering channels are affected.

An ideal testbed to apply this approach of studying quasiparticle interactions upon optical control are charge density waves (CDWs). This ubiquitous symmetry-broken ground state is characterized by a periodic charge- and lattice superstructure that opens up an electronic energy gap at $E_\mathrm{F}$. Photoexcitation allows manipulating the system's underlying potential energy surface (PES), which triggers a transient melting of the CDW, i.e., a photoinduced insulator-to-metal transition~\cite{schmitt2008}, and thus allows studying the impact of the low-energy band structure on the self-energy.

\begin{figure}[t]
\centering
\includegraphics[]{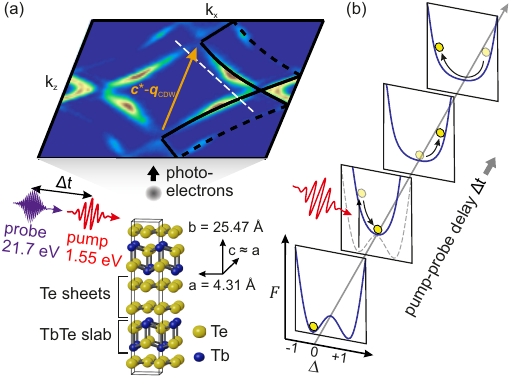}
\caption{Experimental scheme. (a) Top: Symmetrized FS of TbTe$_3$ ($T=100$~K) with tight binding calculations of main (solid) and 3D-backfolded (dashed) bands. Spectral weight along the segments connected by the nesting wave vector $\vec{c}^*\mbox{-}\vec{q}_{\mathrm{CDW}}$ is strongly reduced. The white dashed line marks the momentum cut discussed in Fig.~\ref{fig:exp_data}. Bottom: Quasi-2D atomic structure of TbTe$_3$ consisting of TbTe slabs and Te sheets. By convention, the in-plane crystal directions are along the a and c (k$_x$ and k$_z$) axes. (b) Schematic PES as a function of pump-probe delay after photoexcitation. The yellow circle indicates the system´s order parameter, a metric for the degree of symmetry-breaking of the CDW-to-metal transition, with $|\Delta|=0$ in the metallic and $0<|\Delta|\leq 1$ in the CDW phase. Photoexcitation transforms the ground-state double-well potential into a high-symmetry state, launching an oscillation between metallic and CDW phase.}
\label{fig:schematic}
\end{figure}

Here, we use trARPES to investigate the electron dynamics of a prototypical CDW compound of the rare-earth tritelluride family, TbTe$_3$, after ultrafast near-infrared excitation, as illustrated in Fig.~\ref{fig:schematic}(a). In this material class, strong photoexcitation transforms the ground-state double-well PES to a single-well shape, sketched in Fig.~\ref{fig:schematic}(b). This initiates a melting of the CDW, as the system relaxes towards the new minimum corresponding to the metallic phase. However, for sufficiently strong excitation, the system overshoots across the minimum to the other side of the potential, leading to a reemergence of the CDW, followed by several damped oscillation cycles between metallic and CDW order~\cite{huber2014, beaud2014, trigo2019}, evident from a transient modulation of the CDW energy gap~\cite{maklar2020nonequilibrium}. Besides the collective CDW excitation, we track the photocarrier population of high-energy states and discover a concurrent oscillation of scattering rates at the frequency of the CDW modulation. We find that relaxation of hot photocarriers accelerates when the system is in the metallic phase and slows down when the CDW gap reopens. To understand the underlying microscopic scattering processes, we employ a time-dependent nonequilibrium Green's function formalism (td-NEGF)~\cite{balzer_nonequilibrium_2012,aoki_nonequilibrium_2014, sentef2013examining}, which reveals that the CDW gap opening critically reduces the phase space of e--e scattering. Additional simulations taking into account e--ph coupling demonstrate that phonon scattering remains unaffected by the CDW gap modulation in the examined energy range.

We perform trARPES measurements on single crystals of TbTe$_3$ as described earlier~\cite{maklar2020nonequilibrium}. The setup includes a tabletop fs XUV source ($h\nu_{\mathrm{probe}}=21.7$~eV) with a synchronized optical pump laser ($h\nu_{\mathrm{pump}}=1.55$~eV), using a hemispherical analyzer for photoelectron detection~\cite{puppin2019}. The ultimate time and energy resolutions are $\sim 35$~fs and $\sim 150$~meV, respectively. All measurements were carried out in $p<1$ $\times$ 10$^{-10}$~mbar and at $T=100$~K, well below the transition temperature $T_{\mathrm{c}}=336$~K of the unidirectional CDW phase of TbTe$_3$~\cite{ru2008}. Note that in rare-earth tritellurides strongly wave-vector dependent e--ph coupling~\cite{maschek2015}, in conjunction with a moderately well-nested Fermi surface (FS)~\cite{laverock2005fermi}, leads to a unidirectional CDW in which some parts of the FS become gapped while others remain metallic~\cite{brouet2008}, see Fig.~\ref{fig:schematic}(a).

\begin{figure*}[t!]
\centering
\includegraphics[]{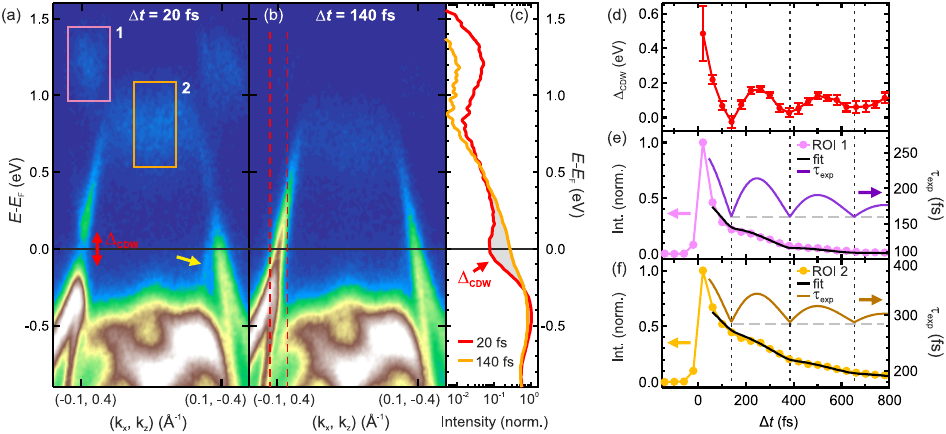}
\caption{Experimental band structure dynamics. (a) Energy-momentum cut along the dashed white line in Fig.~\ref{fig:schematic}(a) at $\Delta t=20$~fs and (b) at $\Delta t=140$~fs for an absorbed fluence of 0.45~mJ/cm$^{2}$. The yellow arrow indicates a replica band arising from the CDW potential. (c) EDCs of the momentum region marked by red lines in panel (b), featuring the CDW gap at $E_\mathrm{F}$ at $\Delta t=20$~fs and subsequent metallization within $\Delta t=140$~fs. (d) Extracted transient CDW energy gap with one standard deviation as uncertainty. Details of the analysis are presented in the Supplemental Material~\cite{supplement}. (e-f) Transient photoemission intensities of the ROIs indicated in panel (a) with a single-exponential decay fit (solid black line) using a time-dependent oscillatory lifetime $\tau_\mathrm{exp}$ (right axis). The data point near $\Delta t=0$~fs is excluded from the fit, as the electrons have not yet reached a thermal distribution, see Supplemental Material~\cite{supplement} for details. The dashed lines serve as guides to the eye.}
\label{fig:exp_data}
\end{figure*}

We investigate the electron dynamics after photoexcitation in the energy-momentum cut shown in Fig.~\ref{fig:exp_data}(a). This allows us to simultaneously track the CDW energy gap $\Delta_\mathrm{CDW}(\Delta t)$ at $E_\mathrm{F}$ (red arrow), the replica bands (yellow arrow),  and the population lifetimes of several distinct high-energy states that are transiently populated upon photoexcitation (regions of interest (ROIs) 1 and 2). As Figs.~\ref{fig:exp_data}(b-c) show, within 140~fs after optical excitation, the system undergoes a transition from the CDW to the metallic phase and the CDW gap closes. Concomitantly, the replica bands vanish, as their spectral weight is transferred back to main bands. However, due to substantial perturbation of the PES, the system subsequently overshoots beyond the metallic phase, followed by several weakly damped oscillations between metallic and CDW order (schematic Fig.~\ref{fig:schematic}(b)), evinced by a transient modulation of the energy gap and replica band intensity~\cite{maklar2020nonequilibrium}.

To quantify the CDW dynamics, we estimate the transient energy gap $\Delta_\mathrm{CDW}$ from the energy distribution curves (EDCs) shown in Fig.~\ref{fig:exp_data}(c), using a leading-trailing edges approach, as discussed in the Supplemental Material~\cite{supplement}. The extracted transient $\Delta_\mathrm{CDW}$ shown in Fig.~\ref{fig:exp_data}(d) features damped oscillations at a frequency far beyond the intrinsic CDW amplitude mode -- a signature of the overshoot regime~\cite{maklar2020nonequilibrium}. Concomitantly, we track the transient photoemission intensities $I$ at $\sim1.2$~eV (ROI 1 in Fig.~\ref{fig:exp_data}(a)) and $\sim0.8$~eV above $E_\mathrm{F}$ (ROI 2), see Figs.~\ref{fig:exp_data}(e-f). Intriguingly, the dynamics of these excited-state populations display a clear correlation with the CDW dynamics, as they feature an exponential decay, on which weak oscillations are imprinted at the frequency of the collective CDW modulation. This oscillatory component directly reflects a modulation of the relaxation rate of the transient population~\cite{energywindows}. In contrast, most previously investigated materials exhibit constant relaxation rates. We quantify the experimental relaxation rates employing a single-exponential decay fit, whereby we use a time-dependent 1/e lifetime $\tau_\mathrm{exp}(\Delta t)$ with a damped sinusoidal contribution to account for the observed modulations~\cite{fullexp}, which yields an excellent description of the experimental data. The deviation from a bare exponential decay is emphasized by the extracted lifetimes $\tau_\mathrm{exp}$, see Figs.~\ref{fig:exp_data}(e-f), as they feature considerable modulations that coincide with the CDW gap dynamics: Whenever the system becomes metallic ($\Delta_\mathrm{CDW}\sim0$\,eV), $\tau_\mathrm{exp}$ reaches its minimum; the high-energy population relaxes faster. Conversely, when the CDW gap opens, the relaxation of the high-energy population slows down, indicated by local maxima of $\tau_\mathrm{exp}$. While it seems natural to assign the observed modulations to a transfer of spectral weight between main bands (captured by the ROIs in Fig.~\ref{fig:exp_data}(a)) and replica bands (yellow arrow), this would yield the opposite effect: A reduction of intensity in the CDW phase due to spectral weight transfer from main to replica bands, and an intensity increase in the metallic phase due to spectral weight transfer back to the main bands~\cite{voit2000}. Hence, a different explanation needs to be invoked.

To elucidate this dynamical modulation of the relaxation, a microscopic perspective onto the scattering channels is required. To this end we employ microscopic simulations based on the td-NEGF formalism, which allows for explicit treatment of e--e interaction and scattering effects beyond mean field.
The relevant bands are captured by the tight binding (TB) model from Ref.~\cite{brouet2008}. We consider the lesser Green's function $G^<_{(m\nu),(m^\prime \nu^\prime)}(\gvec{k}; t,t^\prime) = i \langle \hat{c}^\dagger_{\gvec{k}m^\prime\nu^\prime}(t) \hat{c}_{\gvec{k}m\nu}(t^\prime) \rangle$, where $m, m^\prime$ corresponds to the $p_{x,z}$ orbitals and $\nu=-1,0,+1$ to the nesting index with respect to the CDW wave-vector $\vec{q}_\mathrm{CDW}$. Computing the Green's function provides a direct link to the trARPES intensity~\cite{freericks_theoretical_2009,sentef2013examining}:
\begin{align}
	\label{eq:trarpes_green}
	I(\gvec{k},\omega,\Delta t) \propto \mathrm{Im}\sum_{m} &\int\!d t\!\! \int\! d t^\prime s(t)s(t^\prime) \nonumber \\ &\times e^{i \omega(t-t^\prime)} G^<_{(m0),(m0)}(\gvec{k}; t,t^\prime) \ .
\end{align}
Here, $\omega$ is the binding energy and $\Delta t$ denotes the pump-probe delay, which enters the shape functions $s(t)$. 
While the inequivalence of single-particle and population lifetimes prohibits a direct assignment of experimental relaxation rates to quasiparticle lifetimes~\cite{gierz2014non, yang2015inequivalence, rameau2016energy, baranov_PRB2014}, extracting the population dynamics from the simulated spectra~\eqref{eq:trarpes_green} treats experiment and theory on equal footing and naturally includes population effects and quasiparticle decay~\cite{kemper_mapping_2013,Kemper_PRX_2018}. A subsequent detailed analysis of the self-energy (which determines the Green's function) allows connecting the microscopic quasiparticle picture to the experimentally observed dynamics.
	
\begin{figure}[t!]
\centering
\includegraphics[]{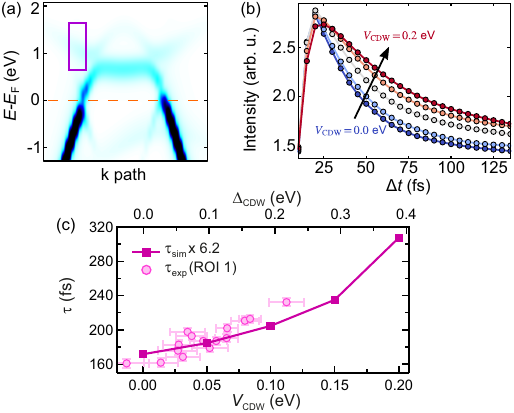}
\caption{td-NEGF simulations. (a) Simulated trARPES spectrum in the CDW state ($V_\mathrm{CDW}=0.2$~eV) after arrival of the pump pulse along the same momentum direction as in the experiment. (b) Intensity integrated over the ROI indicated in (a) as function of $\Delta t$ for varying values of the CDW potential $V_\mathrm{CDW}$. (c) Rescaled simulated lifetimes $\tau_\mathrm{sim}$ versus $V_\mathrm{CDW}$  extracted from (b) using exponential decay fits. For direct comparison to the experiment, the transient lifetimes $\tau_\mathrm{exp}(\Delta t)$ from Fig.~\ref{fig:exp_data}(e) as function of the extracted $\Delta_\mathrm{CDW}(\Delta t)$ from Fig.~\ref{fig:exp_data}(d) are superimposed. The error bars correspond to one standard deviation of the respective fits of $\tau_\mathrm{exp}$ and $\Delta_\mathrm{CDW}$.}
\label{fig:theory_results}
\end{figure}

The model parameter controlling the CDW state is the CDW potential $V_\mathrm{CDW}$ (which corresponds to $\Delta_\mathrm{CDW}\approx 2 V_\mathrm{CDW}$). Varying $V_\mathrm{CDW}$ leads to only minimal changes of the band structure outside of the gapped regions~\cite{voit2000}. Hence, the observed slowdown of the relaxation dynamics must be caused by (dynamical) interaction effects. To test this hypothesis, we include e--e scattering by considering a Hubbard model with weak interaction $U$,
which is supported by the sharpness of the bands and the rather long lifetimes observed in the experiment~\footnote{The Hubbard model serves as a generic model to include e--e scattering. Our main conclusions are equally valid for long-range interactions.}. We treat the e--e self-energy on the level of the second-Born approximation and employ the 
generalized Kadanoff-Baym ansatz (GKBA)~\cite{lipavsky_generalized_1986}. Details are presented in the Supplemental Material~\cite{supplement}.

With these ingredients, we obtain a microscopic description that captures the main features observed in the experiment. However, in contrast to the experiments, we keep the transient gap size constant during relaxation by fixing $V_\mathrm{CDW}$ to highlight the influence of the CDW on the relaxation and to disentangle it from other effects. We simulate the GKBA time evolution upon excitation by a two-cycle pulse with $h \nu_\mathrm{pump}=1.5$~eV and calculate the trARPES signal in the full Brillouin zone~\eqref{eq:trarpes_green}. A typical spectrum along the same path as in Fig.~\ref{fig:exp_data} is shown in Fig.~\ref{fig:theory_results}(a).

Similar to the experimental spectra we have integrated the intensity in the indicated ROI~\cite{energywindows_theory}. The transient photoemission intensity as function of $\Delta t$ is presented in Fig.~\ref{fig:theory_results}(b) for different values of $V_\mathrm{CDW}=0,\dots,0.2$~eV (equidistant steps). For small $\Delta t$ the intensity is only weakly affected by the CDW, while the long-time relaxation exhibits a pronounced dependence on $V_\mathrm{CDW}$. In particular, the decay of intensity in the ROI is significantly faster in the metallic state ($V_\mathrm{CDW}=0$~eV). The data fit well to an exponential decay for $\Delta t > 50$~fs. Extracting the respective lifetime $\tau_\mathrm{sim}$ confirms a monotonic dependence of $\tau_\mathrm{sim}$ on $V_\mathrm{CDW}$, see Fig.~\ref{fig:theory_results}(c). For a quantitative comparison to the experimentally observed dynamical modulation of the lifetime, Fig.~\ref{fig:theory_results}(c) also shows the experimental transient lifetime $\tau_\mathrm{exp}$ of the same ROI as a function of the extracted energy gap $\Delta_\mathrm{CDW}$, which follows a similar monotonic trend. Remarkably, although the absolute simulated values are significantly shorter due to technical constraints~\cite{shortlifetimes}, the relative change of the experimental and simulated lifetimes with $\Delta_\mathrm{CDW}$ is in solid agreement. 

\begin{figure}[t!]
\centering
\includegraphics[]{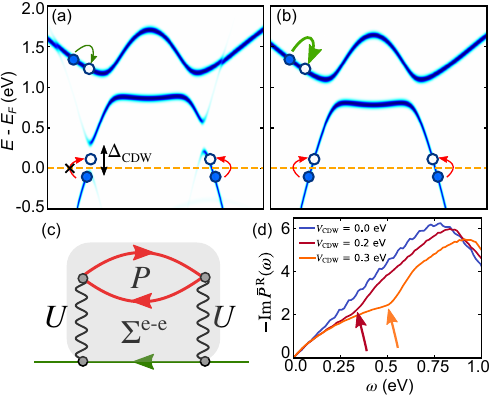}
\caption{Theoretical investigation of e--e scattering. Illustration of the relaxation processes by e--e scattering in (a) the CDW, and (b) in the metallic phase. The presence of a gap $\Delta_\mathrm{CDW}$ in (a) suppresses p--h excitations close to the FS, limiting scattering processes in the upper band. (c) Sketch of the self-energy $\Sigma^{\mathrm{e-e}}$ diagram within the second-Born approximation. The colors are consistent with the scattering processes in (a), (b). (d) Particle-hole polarization averaged over small momentum transfer $|\vec{q}|<0.05$~a.u. for different values of $V_\mathrm{CDW}$.}
\label{fig:theory_schematic}
\end{figure}

The dependence of the lifetime $\tau_\mathrm{sim}$ on $V_\mathrm{CDW}$ explains the superimposed oscillatory component of the relaxation observed in the experiment. Thermalization processes due to e--e scattering in the considered ROI are enhanced in the metallic phase, whereas the presence of the CDW gap renders scattering processes less efficient. To gain intuition on this behavior, we have analyzed the self-energy $\Sigma^{\mathrm{e-e}}$ entering the simulations in detail. In essence, the second-Born approximation captures the relaxation of an excited electron into an unoccupied lower state upon particle-hole (p--h) excitation from a lower-energy to an unoccupied higher state while obeying momentum and energy conservation.
A gap near $E_\mathrm{F}$ suppresses the latter particle-hole transitions, hence reducing the scattering channels of highly excited electrons (Fig.~\ref{fig:theory_schematic}a). Note that the CDW gap at $E_\mathrm{F}$ is only present at momentum points connected by $\vec{q}_\mathrm{CDW}$; p--h excitations at $E_\mathrm{F}$ are still possible in other parts of the Brillouin zone. The relaxation in the upper band is thus not completely suppressed but reduced. In contrast, all p--h channels are available in the metallic phase, which increases the phase space for e--e scattering and thus enhances the relaxation rate (Fig.~\ref{fig:theory_schematic}b). 

This phase-space picture directly enters the Feynman diagrams for $\Sigma^\mathrm{e-e}$ (Fig.~\ref{fig:theory_schematic}c). For an excited electron to lose the energy $\hbar\omega$ and change its momentum by $\vec{q}$, a corresponding p--h excitation obeying energy and momentum conservation is required. The phase-space availability of p--h processes with energy (momentum) transfer $\omega$ ($\vec{q}$) is captured by the p--h susceptibility $P(\vec{q},\omega)$, also termed polarization in the context of many-body methods~\cite{aryasetiawan_gw_1998}, illustrated in Fig.~\ref{fig:theory_schematic}(c). Since small momentum transfer dominates the relaxation dynamics, we focus on the polarization $\bar{P}(\omega)$ averaged over small momenta $\vec{q}$. The scattering phase space of the charges is directly reflected by the retarded component $\bar{P}^\mathrm{R}(\omega)$, presented in Fig.~\ref{fig:theory_schematic}(d). Comparing the metallic and the CDW phase, the CDW gap opening reduces the available phase space for p--h excitations at lower energy transfer $\omega$. Hence, corresponding relaxation processes with energy transfer $\omega$ are increasingly reduced with $V_\mathrm{CDW}$. Consistent with experimental observations, we thus demonstrate how the photoinduced modulation of the CDW gap affects the relaxation dynamics of high-energy photocarriers.

We have also inspected e--ph scattering as a possible explanation for the observed modulation of the relaxation dynamics, calculating the e--ph self-energy $\Sigma^{\mathrm{e-ph}}$ up to second order in e--ph coupling strength assuming a Holstein model. However, there is no noticeable effect of $V_\mathrm{CDW}$ on $\Sigma^{\mathrm{e-ph}}$ in the relevant high-energy ROI~\cite{supplement}. The intuitive explanation is that only e--ph scattering processes close to the CDW gap are affected. Thus, an energy gap opening near $E_\mathrm{F}$ does not influence e--ph scattering of high-energy electrons. Additional GKBA simulations underpin this picture~\cite{supplement}. Hence, we identify e--e scattering as the major relaxation mechanism in the high-energy bands.

In conclusion, we presented a complementary experimental and theoretical study of the electron dynamics in a photoexcited charge-density-wave system, TbTe$_3$. We demonstrated how the CDW state affects the electron self-energy, resulting in a decreased relaxation rate of hot photocarriers, as the energy gap at $E_\mathrm{F}$ critically restricts the phase space of electron-electron scattering. For this, we employed a combined theoretical approach, simulating population lifetimes that allow a quantitative comparison to the experimental photoelectron intensities, and scrutinizing single-particle lifetimes to gain insight into the microscopic details of the interactions, which allows us to link the experimental observations to fundamental interactions. While the examined setting of a collective CDW excitation is rather unique, our conclusions are independent of the details of the CDW mechanism and can be directly transferred to a broad range of materials that feature an insulator-to-metal transition. Furthermore, the array of properties that can be controlled by light is expanding rapidly. Ultrashort optical pulses allow modifying the shape of the Fermi surface~\cite{rettig2016, beaulieu2021ultrafast}, triggering collective modes that are imprinted on the electronic band structure~\cite{de_giovannini_direct_2020, hein2020mode}, and coherently modulating e--ph coupling~\cite{zhang2020coherent}. Hence, the applied approach of tracking the fundamental interactions upon transient optical tuning may facilitate a deeper understanding of a plethora of materials.

\begin{acknowledgments}
The experimental data that support the findings of this study are publicly available~\cite{data2020}.

We thank S. Kubala and M. Krenz (Fritz-Haber-Institut, Berlin) for technical support. This work was funded by the Max Planck Society, the European Research Council (ERC) under the European Union's Horizon 2020 research and innovation program (Grant No. ERC-2015-CoG-682843), the German Research Foundation (DFG) within the Emmy Noether program (Grant No. RE 3977/1 and SE 2558/2), Alexander von Humboldt Foundation (Feodor Lynen scholarship), and the DFG research unit FOR 1700. Crystal growth and characterization at Stanford University (P.W. and I.R.F.) was supported by the Department of Energy, Office of Basic Energy Sciences under Contract No. DE-AC02-76SF00515. 

Y.W.W., L.R., M.P. and C.W.N. carried out the trARPES experiments; P.W. and I.R.F. provided the samples; J.M. analyzed the data; M.Sc. performed the simulations, with guidance from M.Se.; J.M. and M.Sc. wrote the manuscript with support from L.R. and M.Se.; M.W., R.E. and L.R. provided the research infrastructure; all authors commented on the paper.

The authors declare that they have no competing financial interests.
\end{acknowledgments}

\bibliographystyle{apsrev4-2}

\begin{thebibliography}{65}
\bibitem [{ene({\natexlab{a}})}]{energywindows}%
  \BibitemOpen
  \bibinfo {note} {The large energy windows of the ROIs ensure that the total
  respective band intensities are captured during the relaxation. Thus, the
  transient intensity change directly reflects quasiparticle
  scattering.}\BibitemShut {Stop}%
\bibitem [{ene({\natexlab{b}})}]{energywindows_theory}%
  \BibitemOpen
  \bibinfo {note} {The ROI size is chosen large enough to capture the full band
  for any value of $\Delta_\mathrm{CDW}$, thus excluding effects attributed to
  band shifts that could mask the population dynamics.}\BibitemShut {Stop}%
\end{thebibliography}

\end{document}